\documentclass[10pt,conference]{IEEEtran}
\IEEEoverridecommandlockouts
\usepackage{cite}
\usepackage{amsmath,amssymb,amsfonts}
\usepackage{graphicx}
\usepackage{textcomp}
\usepackage{xcolor}
\usepackage{siunitx}
\usepackage{multirow}
\usepackage{colortbl}
\usepackage[ruled]{algorithm2e}
\usepackage{algpseudocode}

\def\BibTeX{{\rm B\kern-.05em{\sc i\kern-.025em b}\kern-.08em
    T\kern-.1667em\lower.7ex\hbox{E}\kern-.125emX}}

\begin{document}

\title{Quantum Support Vector Machine-Based Classification of GPS Signal Reception Conditions

\thanks{This work was supported in part by the Unmanned Vehicles Core Technology Research and Development Program through the National Research Foundation of Korea (NRF) and the Unmanned Vehicle Advanced Research Center (UVARC) funded by the Ministry of Science and ICT (MSIT), Republic of Korea, under Grant 2020M3C1C1A01086407, and in part by the Future Space Navigation and Satellite Research Center through the NRF funded by the MSIT, Republic of Korea, under Grant 2022M1A3C2074404.}
}

\author{\IEEEauthorblockN{Suhui Jeong} 
\IEEEauthorblockA{\textit{School of Integrated Technology} \\
\textit{Yonsei University}\\
Incheon, Korea \\
ssuhui@yonsei.ac.kr} 
\and
\IEEEauthorblockN{Sanghyun Kim} 
\IEEEauthorblockA{\textit{School of Integrated Technology} \\
\textit{Yonsei University}\\
Incheon, Korea \\
sanghyun.kim@yonsei.ac.kr} 
\and
\IEEEauthorblockN{Jiwon Seo${}^{*}$} 
\IEEEauthorblockA{\textit{School of Integrated Technology} \\
\textit{Yonsei University}\\
Incheon, Korea \\
jiwon.seo@yonsei.ac.kr}
{\small${}^{*}$ Corresponding author}
}

\maketitle

\begin{abstract}
Global Positioning System (GPS) plays a critical role in navigation by utilizing satellite signals, but its accuracy in urban environments is often compromised by signal obstructions.
Previous research has categorized GPS reception conditions into line-of-sight (LOS), non-line-of-sight (NLOS), and LOS+NLOS scenarios to enhance accuracy.
This paper introduces a novel approach using quantum support vector machines (QSVM) with a ZZ feature map and fidelity quantum kernel to classify urban GPS signal reception conditions, comparing its performance against classical SVM methods.
While classical SVM has been previously explored for this purpose, our study is the first to apply QSVM to this classification task.
We conducted experiments using datasets from two distinct urban locations to train and evaluate SVM and QSVM models.
Our results demonstrate that QSVM achieves superior classification accuracy compared to classical SVM for urban GPS signal datasets.
Additionally, we emphasize the importance of appropriately scaling raw data when utilizing QSVM.
\end{abstract}

\section{Introduction}

Global positioning system (GPS) is one of the global navigation satellite systems (GNSS) that utilize signals transmitted from satellites to determine users' positions, and GPS is the most commonly used positioning and navigation system.
However, in urban areas, GPS signals are often obstructed or reflected by buildings, resulting in degraded accuracy \cite{Jia21:Ground}.
Previous studies \cite{Kim23:Machine} have attempted to address this issue by classifying GPS signal reception conditions into line-of-sight (LOS), non-line-of-sight (NLOS), and situations where both LOS and NLOS signals are received simultaneously (LOS+NLOS) to enhance positioning accuracy.

Support vector machine (SVM) \cite{Vapnik13:nature} stands out as one of the most effective machine learning algorithms for solving classification problems, particularly through the use of kernel methods.
With ongoing research in quantum computing within the field of machine learning, a novel type of kernel called the quantum kernel has been proposed \cite{Havlivcek19:Supervised}, which offers the potential for quantum advantage unattainable by classical computers.
SVMs using a quantum kernel are referred to as quantum support vector machines (QSVM) \cite{Havlivcek19:Supervised}.
In this paper, we propose a method for classifying urban GPS signal reception conditions using QSVM and compare its classification performance with that of classical SVM.

\section{Experimental Setup}
\label{sec:Experiment}

\subsection{GPS Signal Data}

The GPS signal data were collected using dual-polarized antennas and GNSS receivers from locations P1 and P2, which were also used in our previous study \cite{Kim23:Machine}.
The features for training included the difference in carrier-to-noise density ratio ($C/N_{0}$) between right-hand circularly polarized (RHCP) and left-hand circularly polarized (LHCP) signals, along with the satellite elevation angle.
Table \ref{tab:Data} shows the number of data samples per dataset used in our experiments.
The T0 and T1 datasets are from location P1, while T2 is from location P2.
Due to limitations of the quantum simulator, we used a small number of features and data samples for QSVM experiments.
All feature values were scaled to a range of 0 to 1 using the \textit{MinMaxScaler} from \textit{scikit-learn} \cite{Pedregosa11:Scikit-learn}.

\begin{table}
\centering
 \setlength{\tabcolsep}{15pt}
 \renewcommand{\arraystretch}{1.5}
      \caption{Number of data samples collected from locations P1 and P2}
    \begin{tabular}{c|cc|c}
    \hline
    Location                         & \multicolumn{2}{c|}{\cellcolor[HTML]{EFEFEF}{\color[HTML]{000000} P1}}                                                     & \cellcolor[HTML]{EFEFEF}{\color[HTML]{000000} P2} \\ \hline
    Data set                         & \multicolumn{1}{c|}{\cellcolor[HTML]{EFEFEF}{\color[HTML]{000000} T0}} & \cellcolor[HTML]{EFEFEF}{\color[HTML]{000000} T1} & \cellcolor[HTML]{EFEFEF}{\color[HTML]{000000} T2} \\ \hline\hline
    \cellcolor[HTML]{EFEFEF}Total    & \multicolumn{1}{c|}{160}                                               & 41                                                & \multicolumn{1}{c}{120}                          \\ \hline
    \cellcolor[HTML]{EFEFEF}LOS      & \multicolumn{1}{c|}{80}                                                & 23                                                & 80                                       \\ \hline
    \cellcolor[HTML]{EFEFEF}NLOS     & \multicolumn{1}{c|}{40}                                                & 10                                                & 10                                                \\ \hline
    \cellcolor[HTML]{EFEFEF}LOS+NLOS & \multicolumn{1}{c|}{32}                                                & 8                                                 & 30                                                \\ \hline
    \end{tabular}
    \label{tab:Data}
\end{table}

\subsection{Experimental Configuration}

The experiments were conducted in two phases.
Initially, SVM and QSVM models were trained using the T0 dataset and tested on the T1 dataset.
Subsequently, SVM and QSVM models were trained with the combined T0 and T1 datasets and evaluated on the T2 dataset.

\begin{figure}
    \centering
    \includegraphics[width=1\linewidth]{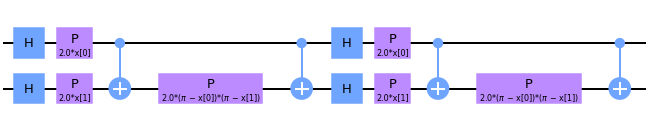}
    \caption{Qauntum circuit for the ZZ feature map.}
    \label{fig:ZZFeaturemapCircuit}
\end{figure}

The QSVM model used in this study employed the ZZ feature map as the feature map and utilized a quantum kernel that calculates fidelity using the compute-uncompute method \cite{Havlivcek19:Supervised}.
The ZZ feature map quantum feature map that initializes the quantum circuit's initial state to all $\left|0\right>$ states, then applies quantum gate operations using the data as parameters for the unitary gates. 
Fig. \ref{fig:ZZFeaturemapCircuit} illustrates the quantum circuit of ZZ feature map.
After the feature mapping of data, the fidelity quantum kernel calculates the fidelity between quantum states. 
All QSVM experiments were conducted utilizing the Qiskit Sampler primitive rather than actual quantum computers, with a specified shot count of 1000.
For comparison, the classical SVM model used the radial basis function (RBF) kernel. 
Neither the QSVM nor the SVM models underwent parameter optimization in this study.

\section{Results and Discussion}
\label{sec:Result}

\begin{figure}
    \centering
    \includegraphics[width=1\linewidth]{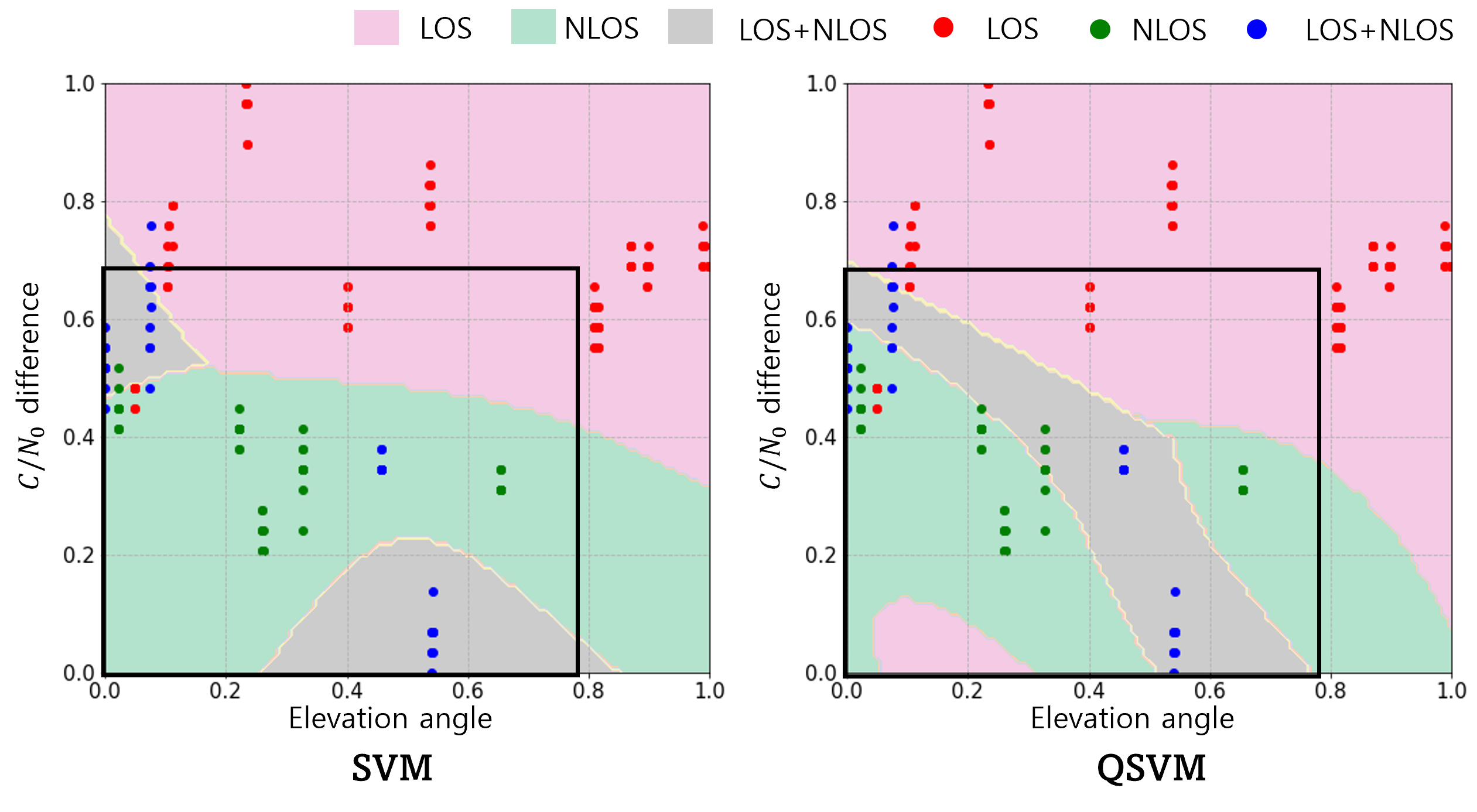}
    \caption{Comparison of decision boundaries between SVM and QSVM trained with the T0+T1 dataset}
    \label{fig:DecisionBoundary}
\end{figure}

Decision boundaries of SVM and QSVM models trained with the T0+T1 datasets are illustrated in Fig. \ref{fig:DecisionBoundary}.
Pink regions denote LOS, green for NLOS, and gray for LOS+NLOS.
Red dots represent LOS data, green dots NLOS, and blue dots LOS+NLOS in T0+T1.
QSVM and SVM exhibit distinct decision boundaries within the marked area. QSVM shows vertically connected LOS+NLOS regions crossing NLOS, while SVM's NLOS region is horizontally connected despite using the same dataset, leading to varied classification accuracy.

The performance validation was conducted using classification accuracy, which is the ratio of correctly classified data points to the total number of data points in the dataset.
Table \ref{tab:Acc} presents the results of evaluating the models trained on the T0 dataset using the T1 dataset, along with the results of training on both the T0 and T1 datasets and evaluating on the T2 dataset.
For the evaluation with the T1 dataset, experimental results showed that the classification accuracy of QSVM reached 0.8, which is higher than the 0.7 accuracy achieved by SVM. 
Regarding the evaluation with the T2 dataset, QSVM achieved a classification accuracy of 0.78, which was higher than the 0.68 accuracy achieved by SVM.
Both QSVM and SVM showed decreased accuracy when evaluated with the T2 dataset compared to T1, as expected due to data collection from different locations.

We investigated the impact of data scaling on QSVM classification accuracy by conducting experiments with raw and scaled datasets. When using raw data, QSVM achieved lower accuracy: 0.61 for training on T0 dataset followed by T1 dataset evaluation and 0.43 for training on combined T0+T1 datasets followed by T2 dataset evaluation, compared to 0.8 and 0.78 respectively with scaled data. This discrepancy arises from the ZZ feature map's sensitivity to data representation, where scaling raw data to a range of 0 to 1 optimizes performance by aligning data within the necessary $0$ to $2\pi$ range for effective quantum state mapping.

\begin{table}
 \setlength{\tabcolsep}{20pt}
    \renewcommand{\arraystretch}{1.5}
    \centering
    \caption{Classification accuracy of SVM and QSVM for the T1 and T2 datasets}
\begin{tabular}{|
>{\columncolor[HTML]{FFFFFF}}c |c|c|}
\hline
     & T1           & T2            \\ \hline
QSVM & \textbf{0.8} & \textbf{0.78} \\ \hline
SVM  & 0.7          & 0.68          \\ \hline
\end{tabular}
\label{tab:Acc}
\end{table}

\section{Conclusion}
\label{sec:Conclusion}

This paper explores the application of QSVM using a ZZ feature map and fidelity quantum kernel for classifying GPS signal reception conditions, contrasting its effectiveness with classical SVM employing an RBF kernel.
The study utilizes T0 and T1 datasets from location P1, and T2 dataset from location P2, with experiments conducted in two phases: training on T0 dataset followed by T1 dataset evaluation, and training on combined T0+T1 datasets followed by T2 dataset evaluation.
Data preprocessing involved scaling to range between 0 and 1.
Results indicate QSVM consistently outperformed SVM in both experimental phases.
Additionally, scaling data to a range between 0 and $2\pi$ significantly influenced QSVM's classification accuracy.

\bibliographystyle{IEEEtran}
\bibliography{mybibfile, IUS_publications}

\vspace{12pt}

\end{document}